\documentclass[useAMS,usenatbib]{mn2e}
\usepackage{times}
\usepackage{amsmath}
\usepackage{amssymb}
\usepackage{graphicx}
\usepackage{xspace}
\usepackage{upgreek}
\usepackage{url}
\usepackage{hyperref}
\usepackage{color}
\usepackage[usenames,dvipsnames]{xcolor}
\usepackage[varg]{txfonts}
\usepackage[T1]{fontenc}
\usepackage[normalem]{ulem}
\usepackage{lineno}

\hypersetup{
    colorlinks=true,
    linkcolor=blue,
    urlcolor=blue,
    citecolor=blue
}

\graphicspath{ {images/} }

\newcommand{\mJy}{\mathrm{mJy}\xspace}
\newcommand{\mJybm}{\mathrm{mJy\,beam^{-1}}\xspace}

\newcommand{\uJybm}{\mathrm{\upmu Jy\,beam^{-1}}\xspace}

\newcommand{\aap}{A\&A}
\newcommand{\araa}{ARAA}
\newcommand{\mnras}{MNRAS}

\newcommand{\apjs}{ApJS}
\newcommand{\apj}{ApJ}
\newcommand{\aj}{ApJ}
\newcommand{\nat}{Nature}

\newcommand{\pasp}{PASP}

\title[Radio imaging the wind collision region of Apep]{AU-scale radio imaging of the wind collision region in the brightest and most luminous non-thermal colliding wind binary Apep}

\author[B.~Marcote et al.]{B. Marcote$^{1,}$\thanks{E-mail: \url{marcote@jive.eu}.},
    J.~R.~Callingham$^{2,3}$,
    M.~De~Becker$^{4}$,
    P.~G.~Edwards$^{5}$,
    Y.~Han$^{6}$,
    R.~Schulz$^{3}$,\newauthor
    J.~Stevens$^{5}$,
    P.~G.~Tuthill$^{6}$\\
$^{1}$Joint Institute for VLBI ERIC, Oude Hoogeveensedijk 4, 7991~PD Dwingeloo, The Netherlands\\
$^{2}$Leiden Observatory, Leiden University, PO\,Box 9513, 2300~RA, Leiden, The Netherlands\\
$^{3}$ASTRON, Netherlands Institute for Radio Astronomy, Oude Hoogeveensedijk 4, Dwingeloo, 7991~PD, The Netherlands\\
$^{4}$Space sciences, Technologies and Astrophysics Research (STAR) Institute, University of Li\`ege, Quartier Agora, 19c, All\'ee du 6 Ao\^ut, B5c, B-4000 Sart Tilman, Belgium\\
$^{5}$CSIRO Astronomy and Space Science, Australia Telescope National Facility, PO Box 76, Epping, NSW~1710, Australia\\
$^{6}$Sydney Institute for Astronomy (SIfA), School of Physics, The University of Sydney, NSW~2006, Australia\\
\vspace{-10pt}
}

\begin{document}

\date{Accepted 2020 December 11  /  Received 2020 December 5}

\pagerange{\pageref{firstpage}--\pageref{lastpage}}
\pubyear{2020}

\maketitle

\label{firstpage}

\begin{abstract}
The recently discovered colliding-wind binary (CWB) Apep has been shown to emit luminously from radio to X-rays, with the emission driven by a binary composed of two Wolf-Rayet (WR) stars of one carbon-sequence (WC8) and one nitrogen-sequence (WN4--6b). Mid-infrared imaging revealed a giant spiral dust plume that is reminiscent of a pinwheel nebula but with additional features that suggest Apep is a unique system.
We have conducted observations with the Australian Long Baseline Array to resolve Apep's radio emission on milliarcsecond scales, allowing us to relate the geometry of the wind-collision region to that of the spiral plume. The observed radio emission shows a bow-shaped structure, confirming its origin as a wind-collision region. The shape and orientation of this region is consistent with being originated by the two stars and with being likely dominated by the stronger wind of the WN4--6b star. This shape allowed us to provide a rough estimation of the opening angle of $\sim 150^\circ$ assuming ideal conditions.
The orientation and opening angle of the emission also confirms it as the basis for the spiral dust plume. We also provide estimations for the two stars in the system to milliarcsecond precision. The observed radio emission, one order of magnitude brighter and more luminous than any other known non-thermal radio-emitting CWB, confirms it is produced by an extremely powerful wind collision. Such a powerful wind-collision region is consistent with Apep being a binary composed of two WR stars, so far the first unambiguously confirmed system of its kind.
\end{abstract}


\section{Introduction}\label{sec:intro}

A significant fraction of massive stars are found in binary or higher multiplicity systems \citep{sana2014}, implying the existence of a large reservoir of systems where the powerful winds of these massive stars can collide and produce strong shocks. Such systems are classified as colliding-wind binaries (CWBs), and the region where the two stellar winds collide is often known as wind-collision region (WCR). WCRs are extremely efficient environments to accelerate particles up to relativistic energies \citep{eichler1993}, producing emission from radio to gamma-rays. WCRs are excellent laboratories to study particle acceleration and non-thermal processes since they operate in a unique parameter space to other astrophysical shocks. While the processes operating in WCRs are equivalent to the ones in supernova remnants or interstellar bow-shocks, WCRs allow studies at higher mass, photon, and magnetic energy densities, while keeping the regions simple enough to be well described by current models and theory \citep[e.g.][]{pittard2006,reimer2006}.

The subset of CWBs known to display non-thermal emission are referred to as Particle-Accelerating Colliding-Wind Binaries (PACWBs). The most complete census of PACWBs contains around 40 systems \citep{debecker2013catalogue,debecker2017}, with a large fraction of them involving evolved massive stars characterized by strong, high kinetic power stellar winds. Among these PACWBs, many contain a Wolf-Rayet (WR) star, which are the end point in the life of massive O-type stars, and are characterized by fast line-driven stellar winds ($\gtrsim 10^3~\mathrm{km\ s^{-1}}$) and significantly higher mass-loss rates ($\gtrsim 10^{-5}~\mathrm{M_\odot\ yr^{-1}}$) than their progenitor form. Therefore, the WCRs created by these types of stars are expected to be more powerful than the stereotypical systems composed of OB-type stars.

However, as lifetimes of WR stars \citep[$\sim 10^{5}\ \mathrm{yr}$;][]{meynet2005} are significantly shorter than their hydrogen-burning progenitor phases, most of the known powerful PACWBs are composed of a single WR star and an OB-type companion. It is expected only a few binary systems composed of two WR stars exist in our Galaxy. Only a potential system of this kind has been reported so far: WR~48a \citep{williams2012,zhekov2014}, although such a classification is contentious since the spectra of WR~48a does not possess emission lines indicative of two WR stars.

\begin{figure}
    \includegraphics[width=0.465\textwidth]{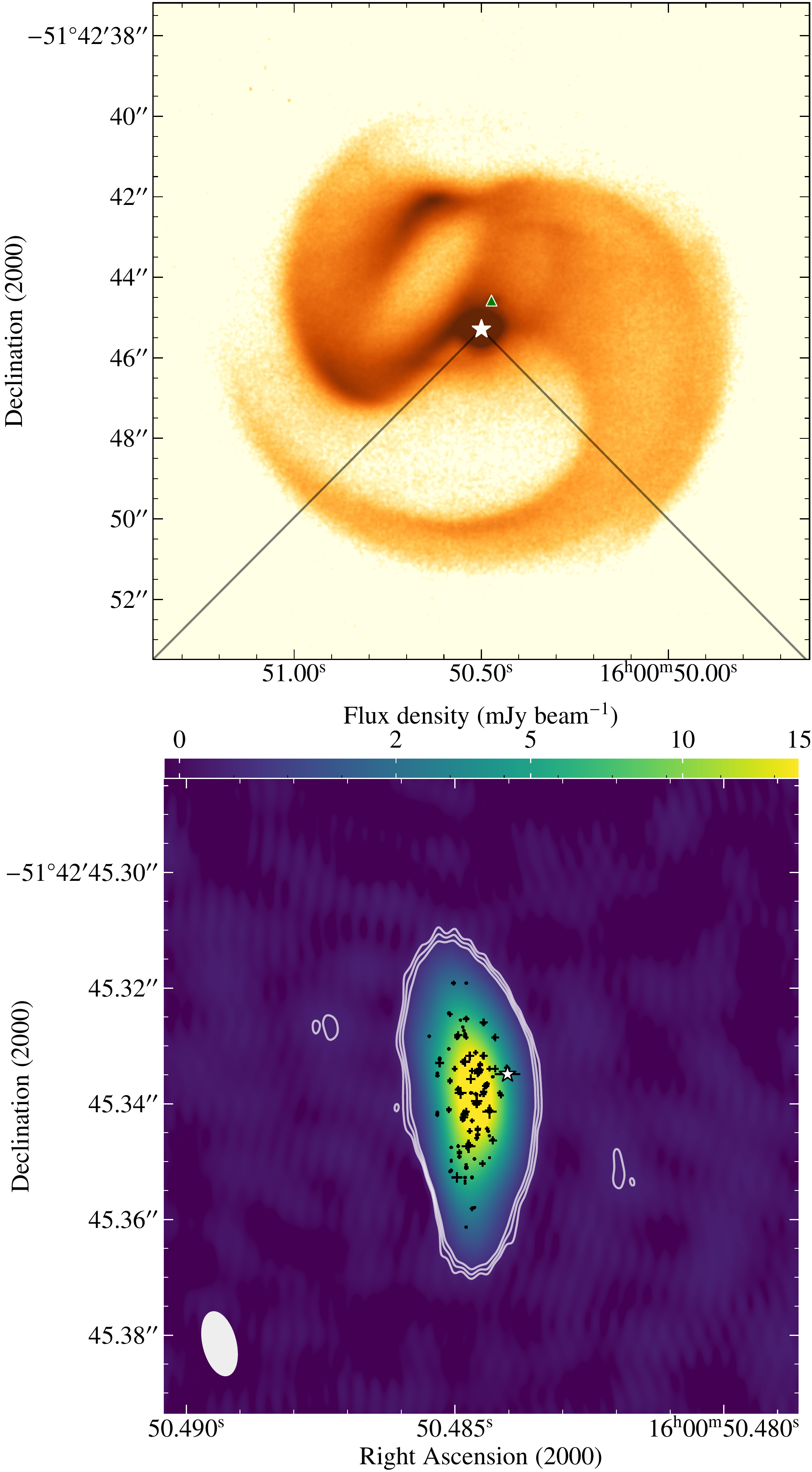}
    \caption{Top panel: Mid-infrared $8.9\text{-}\mathrm{\upmu m}$ image of Apep displaying the dust pattern resembling the pinwheel nebulae but with additional intricate and overall larger structure \citep{callingham2019}. The central white star represents the position of the central engine, while the offset green triangle represents the position of a northern O-type supergiant.
    Bottom panel: Radio image of the central engine obtained by the LBA at 13~cm on 2018 July 18 showing the WCR. Contours start at three times the rms noise level of $38~\uJybm$, and increase by factors of $\sqrt{2}$. Only three contours are shown for clarity. The proper-motion corrected {\em Gaia} DR2 position is shown by the white star, representing the convolution of the position of the WC8 and WN4--6b stars. The black crosses identify the location of the positive-flux clean components that were generated during cleaning (see Sect.~\ref{sec:obs}). Their sizes are proportional to the flux density for each component. The synthesized beam is $11.3 \times 5.6~\mathrm{mas^2},\ \mathrm{PA} = 14^{\circ}$, and is shown at the bottom-left corner. For reference, $10$ and $0.02~\mathrm{arcsec}$ represent $\approx 25\,000$ and $50\ \mathrm{au}$, respectively, at the assumed source distance of $\sim 2.5~\mathrm{kpc}$.}
    \label{fig:lba}
\end{figure}

The source 2XMM~J160050.7$-$514245, hereafter called Apep as in \citet{callingham2019}, was recently discovered as a surprisingly bright radio, infrared, and X-ray emitter \citep{callingham2019}. At the suggested $\approx 2.4\,\mathrm{kpc}$ distance, Apep is the brightest and most luminous non-thermal radio-emitting CWB discovered by over an order of magnitude \citep{callingham2019,debecker2013catalogue}. Mid-infrared images taken by the Very Large Telescope (VLT) revealed an expanding $12\text{-}\mathrm{arcsec}$-diameter spiral dust plume (see top panel of Fig.~\ref{fig:lba}). This structure, similar to the known pinwheel nebula but on a larger scale, are characteristic of CWBs that are composed of at least one carbon-rich WR star \citep{tuthill1999,tuthill2008}, as the physical conditions in the WCR can reach higher densities than usual to boost dust formation \citep{williams2009}. Modeling of the pinwheel nebula and its expansion suggested the presence of a long-period ($\sim 100\ \mathrm{yr}$) CWB \citep{callingham2019}. One peculiarity of Apep inconsistent with standard CWB physics is that the measured expansion velocity of the spiral dust plume is at nearly an order of magnitude slower than that expected from the spectroscopically measured stellar winds. \citet{callingham2019} suggested that one possible explanation for this mismatch of measured velocities is that one of the WR stars in Apep is rapidly-rotating, producing a slow and dense equatorial wind. Such a conclusion would make Apep a potential Galactic progenitor to long-duration gamma-ray bursts \citep[e.g.][]{cantiello2007}.

\citet{callingham2019} predicted the existence of an unresolved massive binary at the centre of Apep, and resolved a potential companion O-type supergiant $\approx 0.7\text{-}\mathrm{arcsec}$ away from the central binary. Recent observations with the X-SHOOTER spectrograph on the VLT have confirmed the existence of two classical Wolf-Rayet stars of one carbon-sequence (WC8) and one nitrogen-sequence (WN4--6b) in the central binary, with terminal line-of-sight wind velocities of $v_{\infty,{\rm WC}} = 2\,100 \pm 200\ \mathrm{km\ s^{-1}}$ and $v_{\infty,{\rm WN}} = 3\,500 \pm 100\ \mathrm{km\ s^{-1}}$, respectively \citep{callingham2020}.
Additionally, through the use of aperture masks with the NACO camera on the VLT, \citet{han2020} have constrained the separation between the two WR stars to be consistently $D = 47 \pm 5\ \mathrm{mas}$, with a position angle of ${\rm PA} = 273 \pm 2^\circ$ (2016 April 28) and $278 \pm 3^\circ$ (2019 March 21--24). By model fitting the dust structure, a distance of $\sim 2.5\ \mathrm{kpc}$ was also suggested for the system.

However, one of the remaining mysteries of Apep is how the unusually bright non-thermal synchrotron radio emission is being driven and if this is consistent with the standard CWB model \citep{callingham2019}. Radio emission in CWBs is typically dominated by the WCR near the stagnation point, and many systems have previously been well described by current models \citep[see e.g.][]{dougherty2003,pittard2006}. Very high resolution radio observations using Very Long Baseline Interferometry (VLBI) have been successful in resolving WCRs, confirming orbital properties, and determining the physical properties of the binary systems and their individual stars \citep[see e.g.][]{dougherty2005,benaglia2015}.

To understand the nature of Apep and its radio emission, we conducted VLBI radio observations with the Southern Hemisphere Long Baseline Array \citep[LBA;][]{edwards2015} that allowed us to resolve the emission associated with the wind-collision region.
In Section~\ref{sec:obs} we describe these observations and the data reduction. Section~\ref{sec:results} details the obtained results and the modelling of the radio-emitting region. The implications of these results for understanding the dynamics of Apep are discussed in Section~\ref{sec:discussion}. We summarise the conclusions of this paper in Section~\ref{sec:conclusions}.

\section{Observations and data reduction}\label{sec:obs}

Apep was observed with the LBA at 13~cm (2.28~GHz central frequency) on 2018 July 18 from 05:00 to 17:00~UTC (project code V565). Ten stations participated in this observation: the Australian Telescope Compact Array (ATCA), Ceduna, Hartebeesthoek, Hobart, Katherine, Mopra, Parkes, Tidbindilla, Warkworth, and Yarragadee.
The data were recorded with a total bandwidth of 64~MHz and divided during correlation with the DiFX software correlator \citep{deller2011} into four subbands of 32 channels each. Most stations recorded full polarization, except Parkes, Tidbindilla and Warkworth, that only recorded right circular polarization.

The source PKS~B1622$-$297 (J1626$-$2951) was used as fringe finder and bandpass calibrator. PKS~B1934$-$638 was used for phasing-up ATCA. PMN~J1603$-$4904 (J1603$-$4904 hereafter; located $2.7^{\circ}$ away from Apep) was used as phase calibrator in a phase-referencing cycle of 3.5~min on target and 1.5~min on the calibrator. As a result, Apep was observed for a total of  $\approx 7.2\ \mathrm{h}$.

The LBA data have been reduced in {\sc AIPS}\footnote{The Astronomical Image Processing System ({\sc AIPS}) is a software package produced and maintained by the National Radio Astronomy Observatory (NRAO).} \citep{greisen2003} and {\sc Difmap} \citep{shepherd1994} following standard procedures. \emph{A-priori} amplitude calibration was performed using the known gain curves and system temperature measurements when recorded on each station during the observation. Nominal System Equivalent Flux Density (SEFD) values were used for Ceduna, Hobart, Katherine, Tidbindilla, Warkwork, and Yarragadee.
We manually removed bad data: missing polarizations, bad subbands, or times and frequencies affected by radio frequency interference. We first corrected for the instrumental delays and bandpass calibration using J1626$-$2951, and thereafter fringe-fit the data using all calibrator sources. The phase calibrator was then imaged and self-calibrated to improve the final calibration of the data. The obtained solutions were transferred to Apep, which was finally imaged using the \textsc{clean} algorithm \citep{clark1980}.

\section{Results on Apep's radio emission} \label{sec:results}

Apep is detected on milliarcsecond scales from the LBA data as a bow-shaped radio source with a peak brightness of $20.54 \pm 0.04~\mJybm$ and a total flux density of $58.6 \pm 1.3~\mJy$ at 13~cm (see bottom panel of Fig.~\ref{fig:lba}). The emission is spread over a region of $\sim 60 \times 30\ \mathrm{mas^2}$ with a centroid, inferred from a Gaussian fitting to the radio image, located at the (J2000) position $\alpha_{\rm WCR} = 16^{\rm h}0^{\rm m}50.48467^{\rm s} \pm 0.3\ \mathrm{mas},\ \delta_{\rm WCR} = -51^\circ 42^\prime 45.3385^{\prime\prime} \pm 0.5\ \mathrm{mas}$. The quoted uncertainties are composed of the statistical uncertainties ($0.16$ and $0.3\ \mathrm{mas}$ for $\alpha$ and $\delta$, respectively), the uncertainties in the absolute International Celestial Reference Frame position of the phase calibrator \citep[$0.26~\mathrm{mas}$;][]{beasley2002,gordon2016}, and the estimated uncertainties from the phase-referencing technique \citep[$0.06$ and $0.19\ \mathrm{mas}$;][]{pradel2006} added in quadrature. No additional sources of significant ($>6\sigma$) radio emission are reported in an $6 \times 6\text{-}\mathrm{arcsec^2}$ area centred on Apep above the rms noise level of $40\ \mathrm{\upmu Jy}$.

The obtained position for the radio emission lies close to the position provided by the {\em Gaia} data release 2 (DR2) Catalog \citep{gaia2016,gaia2018} for the central binary (with source ID 5981635832593607040) after accounting for proper motion (see bottom panel of Fig.~\ref{fig:lba}). The separation between both positions ($\approx 7\ \mathrm{mas}$) represents a $< 2$-$\sigma$ offset after accounting for uncertainties. In any case, we note that the {\em Gaia} position is composed of optical emission arising from both stars of the central binary, which remain unresolved in their data. The radio and optical positions are thus expected to be close but not fully agree, and it is unclear how the binary motion of the stars impacts the precision of the {\em Gaia} DR2 position\footnote{The parallax reported in the {\em Gaia} DR2 Catalog is labelled as not reliable as it can be seen by the large values of the astrometric goodness of fit and astrometric excess noise. \citet[Sect.~1.2.1 in the supplementary information]{callingham2019} provided a possible explanation for the large uncertainties for this system: the elongated pixels from {\em Gaia} only resolve the emission from the central binary and the northern O-type star for some orientations. This can thus bias the parallax measurements.}.

The reported radio flux density from the LBA data is roughly a factor of two lower ($\sim 60$ against $\sim 120\ \mathrm{mJy}$) than the flux density measured from ATCA data \citep{callingham2019}, where the source remains unresolved on $\sim 0.1~\mathrm{arcsec}$ scales. While the absolute amplitude calibration of the LBA data can exhibit uncertainties up to $\sim 20\%$ due to the intrinsic calibration procedures related to VLBI arrays, the observed difference is too large to be explained solely by standard amplitude calibration uncertainties.
The missing flux is also unlikely to be related to intrinsic source variability. While the radio emission is known to be variable \citep{callingham2019}, such variability is only observed on very long (several year) timescales.
However, we note that Apep is significantly resolved by the LBA data, which is sensitive to milliarcsecond scales. This is clear from both the image plane, where the source shows a size significantly larger than the synthesized beam by roughly up to a factor of five (see Fig.~\ref{fig:lba}), and from the $(u,v)$-plane, where the longest baselines of the LBA did not detect any significant emission. We note that the LBA lacks short baselines (the shortest baselines are 114 and 207~km, from ATCA--Mopra and Mopra--Parkes, respectively). Therefore, it is expected that a significant fraction of the radio emission is resolved out in the LBA data. The compactness of the emission in the ATCA data guarantees that all the radio emission reported for Apep belongs to the WCR \citep{callingham2019}.

It is important to note that the missing flux density likely represents only a tiny fraction of the total peak brightness in the LBA image, given the small synthesized beam relative to the source size. We can consider that the missing $\sim 60\text{-}\mathrm{mJy}$ flux density is spread over those angular scales that provide the most extended significant emission of the LBA image (lowest contours in Fig.~\ref{fig:fit}), which cover an area of about $5 \times 5$ times the synthesized beam. In this case, such missed emission would only exhibit a peak brightness of $\sim 2\ \mathrm{mJy\ beam^{-1}}$, which represents only $\sim 10\%$ of the observed peak brightness. In a more realistic scenario where at least part of such missed emission arises from a more extended region, the associated peak brightness would be even lower. We thus conclude that the observed radio region is a trustworthy representation of the predominant location of radio emission in Apep.

Additionally, the lack of compact radio emission outside the WCR (above the rms noise level of $40\ \mathrm{\upmu Jy}$) also implies that no significant interaction occurs between the central binary and the northern O-type supergiant. We also clarify that the individual stellar winds of each of the three stars are not detected. Their individual thermal emission is estimated to lie below the 0.1-mJy level at 13~cm, using the formalism proposed by \citet{wright1975}. At a distance of about 2.5~kpc, such winds would be more likely detected at shorter wavelengths, considering the thermal nature of the emission.

\section{Discussion} \label{sec:discussion}

The observed bow-shaped radio emission is representative of a WCR. However, the reported radio flux densities make Apep brightest and most luminous radio-bright CWB discovered to date. The presence of two WR stars, both exhibiting powerful winds, may introduce a significant difference in the properties involved at the WCR with respect to the more common CWBs comprising one WR star and an OB star. For example, the emission could be bolstered by a significantly higher kinetic power budget for the winds, and/or stronger magnetic fields in the shock than typical. A larger energy budget and magnetic field strength could go some way to explaining why Apep displays synchrotron radio emission one order of magnitude brighter and more luminous than other CWBs. We note that the observed synchrotron emission is expected to be proportional to the magnetic field as $B^{3/2}$ and to the electron density \citep{longair2011}, where the latter is related to $\dot M\, v^{-1}$ \citep{canto1996}. The significantly slower winds and higher mass-loss rates of the WC star could then bolster the radio emission.

Furthermore, the separation between the two stars may be optimal for luminous non-thermal radio emission: being close enough to produce a powerful shock, but far enough to avoid free-free absorption at gigahertz-frequencies. The size of the radio photosphere (radius of a sphere at which the optical depth is equal to one) of each individual wind can be estimated according to the approach adopted for instance by \citet{debecker2019}. This quantity allows for a rough estimate of the capability of a given stellar wind to attenuate radio emission due to free-free absorption. At 13~cm, one obtains values shorter than $10\ \mathrm{au}$. Assuming the WCR is located roughly midway between the two stars, the projected separation between each star and the stagnation point at $\sim 2.5\ \mathrm{kpc}$ is about $ \sim 59\ \mathrm{au}$. The synchrotron emitting region is thus clearly away from the denser parts of the winds and is thus not affected by free-free absorption.
This is consistent with the absence of signatures of absorption have been reported to date at gigahertz frequencies \citep{callingham2019}, and the observed radio morphology, that would be expected to be centred on the stagnation point of the shock, to occur at some intermediate position between the stars.

For comparison, we note that $\upeta$ Carinae, the most powerful CWB, only shows thermal radio emission due to its closer binary semi-major axis of $\approx 15.4\ \mathrm{au}$ \citep{madura2012}. The putative non-thermal component of $\upeta$ Carinae's WCR is likely to be completely free-free absorbed due to the aforementioned conditions \citep{debecker2013catalogue}.

\subsection{Orientation of Apep's WCR} \label{sec:radioVSoptical}

The central engine in Apep is known to host a WC8 and WN4--6b star \citep{callingham2020}.
The separation between the two stars has been measured by near-infrared NACO data, taken in 2016 April 28 and 2019 March 21--24, to be consistently $D = 47 \pm 5\ \mathrm{mas}$, with position angles of $\mathrm{PA} = 274 \pm 2$ and $278 \pm 3^\circ$, respectively \citep{han2020}. However, the absolute positions of the two stars remained unknown from these data. We note that the orbital period of the central binary is estimated to be $\sim 100\ \mathrm{yr}$, and thus we do not expect significant changes ($\sim 1\%$) in the orbital parameters between any of these epochs.

Since the LBA epoch (2018 July 18) is interleaved with the NACO epochs, we were able to estimate the PA of the system at the LBA epoch to be $277 \pm 3^\circ$. This value for the PA is consistent with the orientation observed in the curved radio emission of Apep (see Fig.~\ref{fig:lba}). We can thus confirm that the observed radio emission arises from the interaction of only these two stars.

A Gaussian fitting to the region revealed that while the radio emission is clearly elongated in the North-South direction (with a semi major axis --- half width at half maximum --- of $\sim 18~\mathrm{mas}$), its extension in the direction between the two stars (roughly East-West) is comparable to the size of the synthesized beam ($\sim 6~\mathrm{mas}$), and thus not significant.
We note that given the scales of the system, the shock must be adiabatic \citep[see e.g.][]{stevens1992}. In this case we would expect the radio emission to encompass the two shock fronts produced by the collision of the two winds, as both would contribute to the particle acceleration, and remaining close to the stagnation point \citep[as reported in systems like WR~140;][]{pittard2006}. The fact that we cannot claim any significant extension along the axis between the two stars imposes an upper-limit to the separation between these two shocks of $\lesssim 12~\mathrm{mas}$ (or $\lesssim 30~\mathrm{au}$ at the assumed distance of $\sim 2.5~\mathrm{kpc}$).

\subsection{A first estimation of the Contact Discontinuity}\label{sec:estimating}

A full understanding of the observed WCR is only possible by conducting a full magnetohydrodynamic and radiative-transfer simulation of the collision between the two stellar winds, that go beyond the purpose of this manuscript.
However, we note that even a simplistic scenario assuming two ideal (spherically symmetric) stellar winds can be typically taken as a first (rough) approach to characterize the observed broad picture of the radio emission, as successfully demonstrated in other CWBs \citep[see e.g.][]{blomme2010,benaglia2015}, and test its consistency with respect to the properties derived from the IR data \citep{callingham2019,callingham2020,han2020}.

\begin{figure}
    \includegraphics[width=0.484\textwidth]{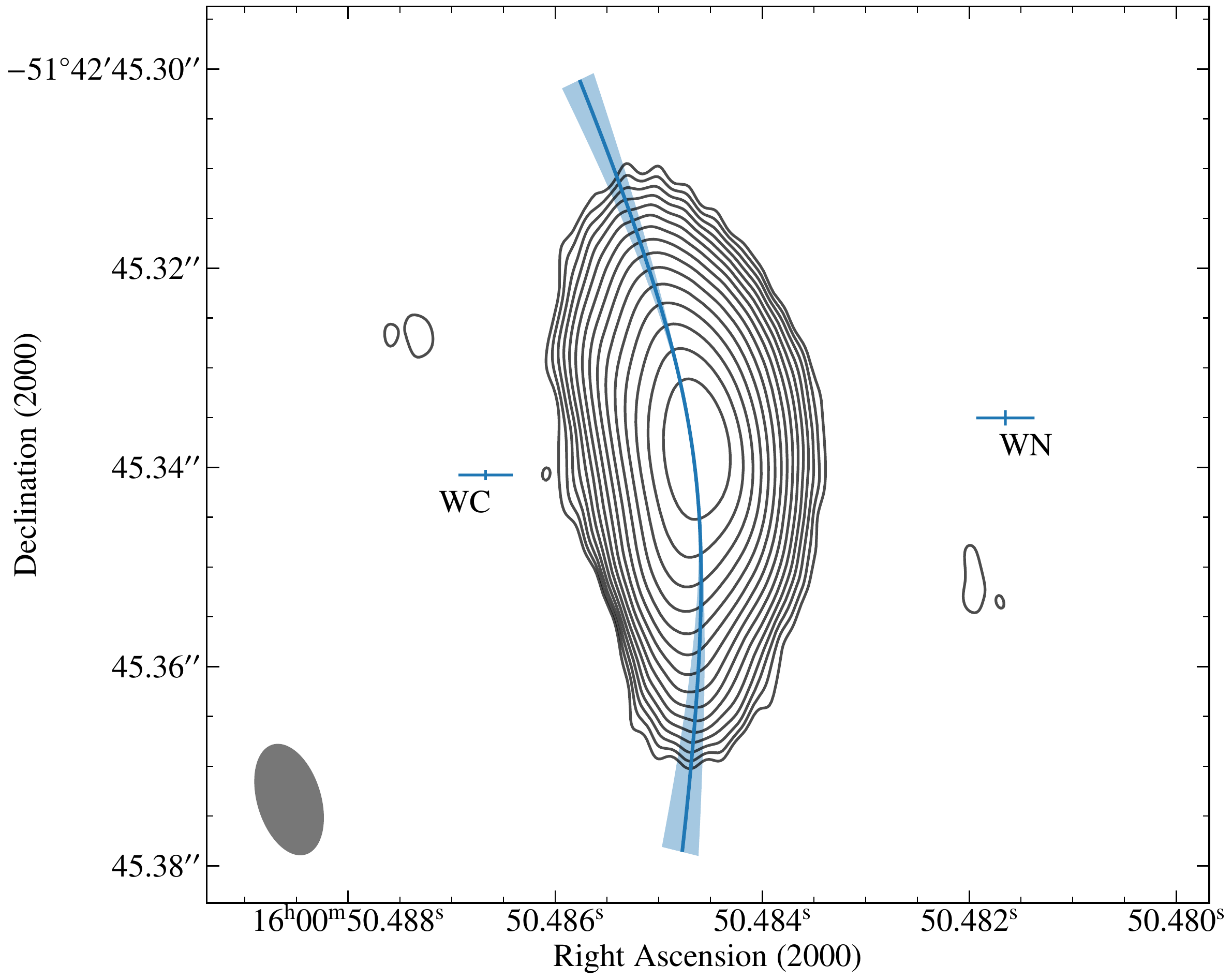}
    \caption{Ideal contact discontinuity (CD) compared to the shape of Apep's radio emission. Contours start at three times the rms noise level of $38~\uJybm$, and increase by factors of $\sqrt{2}$, and represent the same image presented in the bottom panel of Fig.~\ref{fig:lba}. The solid blue line and the shadowed light blue region represent the ideal CD shape for the mean value and $1\text{-}\sigma$ confidence interval of $\eta = 0.44 \pm 0.08$, respectively. The predicted positions for the two stars (WC and WN) from the given CD are represented by the blue crosses, where uncertainties on position (relative to the position of the stagnation point) represent the $1\text{-}\sigma$ confidence interval. The synthesized beam is $11.3 \times 5.6~\mathrm{mas^2},\ \mathrm{PA} = 14^{\circ}$, and is shown at the bottom-left corner. For reference, $0.02~\mathrm{arcsec}$ represent $50\ \mathrm{au}$ at the assumed source distance of $\sim 2.5\ \mathrm{kpc}$.}
    \label{fig:fit}
\end{figure}

Following this path, and following the details provided in Appendix~\ref{sec:app}, one could compare the observed bow-shaped radio emission to the ideal contact discontinuity (CD) shape expected under such scenario. Figure~\ref{fig:fit} shows this ideal CD with respect to the observed emission, suggesting that the observed curvature can still be understood under ideal conditions -- and for the given resolution.

The ideal CD shape provides both a rough estimation of the wind momentum rate ratio -- under the assumption of spherically symmetric and homogeneous winds -- of $\eta = 0.44 \pm 0.08$ (see equation~\ref{eq:eta}) and of the putative positions of the two stars (see Appendix~\ref{sec:app} for details) at the epoch of the LBA observations (2018 July 18):
\begin{align}
	\alpha_{\rm WC} &= 16^{\rm h}0^{\rm m}50.4867^{\rm s}\ \pm 5\ \mathrm{mas},\nonumber\\
	\delta_{\rm WC} &= -51^\circ 42^\prime 45.3408^{\prime\prime}\ \pm 1.6\ \mathrm{mas}, \label{eq:posWC}
\end{align}
\noindent for the WC8 star, and
\begin{align}
	\alpha_{\rm WN} &= 16^{\rm h}0^{\rm m}50.4817^{\rm s}\ \pm 5\ \mathrm{mas},\nonumber\\
	\delta_{\rm WN} &= -51^\circ 42^\prime 45.3350^{\prime\prime}\ \pm 2.0\ \mathrm{mas}, \label{eq:posWN}
\end{align} for the WN4--6b star.
We note that the quoted uncertainties take into account both the statistical uncertainties of the radio data and the curvature on the ideal CD shape, and the systematic uncertainties for the absolute position (not shown in Fig.~\ref{fig:fit}).

The ideal CD shape shown in Fig.~\ref{fig:fit} would also imply an opening angle of $2\theta_w = 150 \pm 9^\circ$, which is in fact consistent with the value of $\sim 150^\circ$ estimated from optical/near-infrared spectroscopic observations \citep{callingham2020}, and slightly higher than the value of $\sim 125^\circ$ determined from the model of the dust expansion \citep{callingham2019,han2020}.
We note that, in any case, we would expect the opening angle from the WCR to deviate from the later one for several reasons, not just because of the ideal conditions assumed in the WCR.
First, it could be expected that the opening angles of the shocks have not reached yet their asymptotic values at the region where the radio emission is produced \citep[see e.g.][]{pittard2006}, but such value is reached at the scales where the dust structure is observed.
Secondly, while the radio WCR is created and dominated by the direct collision of the two stellar winds, the dust plume is more sensitive to the significant clumpiness and turbulence of the winds of the WR stars \citep[see e.g.][]{crowther2007} and of the region with the mixed, shocked, winds. Furthermore, a larger opening angle for the WCR than that derived from the dust plume could actually be expected due to the conditions necessary for dust formation to occur \citep{tuthill2008}. In summary, only a small portion of the winds are expected to collide at or near the stagnation point (the radio-emitting WCR). Most of the wind interactions take place at larger distances along the spiral plume \citep{tuthill2008}, which already deviate significantly from any ideal wind-interaction scenario. This would explain why the typically observed WCR, including this one in Apep, can be consistently explained by such ideal scenarios with no significant deviations, while the full modelling of the system (including the dust plume) would require a more elaborated scenario.

Finally, we note that the assumed ideal CD -- assuming ideal and spherical winds -- would imply a mass-loss rate ratio of $\dot M_{\rm WC}/\dot M_{\rm WN} = 0.73 \pm 0.15$ (considering the measured terminal line-of-sight wind velocities of the two WR stars from spectroscopic observations, \citealt{callingham2020}, of $v_{\infty,{\rm WC}} = 2\,100 \pm 200\ \mathrm{km\ s^{-1}}$ and $v_{\infty,{\rm WN}} = 3\,500 \pm 100\ \mathrm{km\ s^{-1}}$). This value is actually consistent with the typical mass loss rates expected for these kinds of stars:
Galactic WC8 stars show an average mass loss rate of $\dot M_{\rm WC8} \sim 10^{-4.5}\ \mathrm{M_\odot\ yr^{-1}}$ \citep{sander2019}, while WN4--6 stars exhibit mass loss rates ranging $10^{-4.8}$ to $10^{-3.8}\ \mathrm{M_\odot\ yr^{-1}}$ \citep{hamann2019}. We can then adopt an average value of $\dot M_{\rm WN4\text{--}6b} \sim 10^{-4.3}\ \mathrm{M_\odot\ yr^{-1}}$, which would imply a ratio of $\sim 0.63$, consistent with the aforementioned one.

However, the expansion of the dust plume has been measured to be significantly slower ($\sim 600\ \mathrm{km\ s^{-1}}$) than the $\sim 2\,500\ \mathrm{km\ s^{-1}}$ the dust is expected to inherit from the collision of the winds with the aforementioned mass loss rates and terminal wind speeds \citep{callingham2020}. One possible solution to this discrepancy is that the wind of at least one of the stars, likely the WC8, is highly asymmetric, with a putative dense, slow equatorial wind potentially produced by a rapid rotation of the star \citep{callingham2019,callingham2020,han2020}.
In this case, one could expect its stellar wind to be as slow as $\lesssim 1\,000\ \mathrm{km\ s^{-1}}$ to explain the dust expansion. In such a case, that would imply a mass-loss rate ratio of $\dot M_{\rm WC8}/\dot M_{\rm WN4\text{--}6b} \sim 1.5$. Assuming that the WC8 star still exhibits a typical mass-loss rate, we estimate $\dot M_{\rm WN4\text{--}6b} \sim 10^{-4.7}\ \mathrm{M_\odot\ yr^{-1}}$, which still lies within the observed range of mass-loss rates for these types of stars \citep{hamann2019}. Assuming, on the other hand, an average mass-loss rate for the WN4--6b star, we estimate $\dot M_{\rm WC8} \sim 10^{-4.1}\ \mathrm{M_\odot\ yr^{-1}}$, which would imply a value lower than the ones observed for WC8 stars \citep{sander2019}. However, we remark that this value would be an estimation for the mass-loss rate of the WC8 star only valid at equatorial latitudes, as we are under the assumption of non-spherically-symmetric winds.

The two extreme cases (symmetric or highly-asymmetric winds) provide then consistent results with the observed curvature of the WCR. The existence of a more likely scenario where at least one of the winds may exhibit a inhomogeneous profile would imply that the wind momentum rate ratio estimation must be redefined by taking into account such profile (not fully understood to date) and the inclinations of the two star spins with respect to the orbital plane.\\

To summarize, we have imaged the radio emission of Apep, whose structure is consistent with a WCR produced by the WC8 and WN4--6b stars, and thus directly resolves the nature of Apep as a CWB. The orientation, position, and opening angle of the observed bow-shaped structure are consistent with the positions derived from the NACO \citep{han2020} and X-SHOOTER data \citep{callingham2020}. Additionally, even a scenario assuming two spherical winds with ideal conditions provide a consistent description of the system.

Following VLBI observations of the source with a cadence of years would allow us to reconstruct the orbital motion of the two stars and the evolution of the WCR. These data would unambiguously and accurately constrain the orbit of Apep. These data, together with the spectral information from hundreds of megahertz to tens of gigahertz, would allow us to fully characterize the radio-emitting region; constraining not only the dynamic properties of the winds but also the particle density and magnetic field at the WCR. In particular, it would be interesting whether, or at what phase of the orbit, the WCR becomes fully free-free absorbed. Such a measurement would provide another independent estimate of the mass-loss rate present in the system \citep{dougherty2003}. In general, we would expect these additional parameters would help reveal the fundamental differences in Apep with respect to other CWBs to explain why radio emission is so bright.

\section{Conclusions} \label{sec:conclusions}

The LBA data presented here allowed us to resolve the radio emission of Apep, revealing the presence of a strong wind collision region produced by the WC8 and WN4--6b stars.  The observed orientation and opening angle of such bow-shock region are consistent with the dust spiral structure observed by \citet{callingham2019} and modelled by \citet{han2020}. We confirm that the full structure (radio-emitting WCR and dust pinwheel nebula) are produced by these two WR stars in the same shock and no additional interactions are required from third components, like the nearby O-type supergiant of the system. Apep is the first known particle-accelerating colliding-wind binary unambiguously composed of two WR stars, and is the brightest and most luminous PACWB by an order of magnitude \citep{debecker2013catalogue}.

Apep thus belongs to a select group of binaries composed of two WR stars. As the WR is a short-lived stage in the life of stars, it is expected that this kind of binaries are rare. Apep is thus a valuable object that allows us to study in detail these extreme wind interactions and the evolution of these types of stars.
Given that these systems are potential progenitors of long gamma-ray bursts \citep[e.g.][]{cantiello2007}, the resulting studies would provide valuable data to constrain the evolution and dynamics of these systems prior to cataclysm.

To finalize, we have shown how VLBI observations have been once again a successful approach to unveil the nature of CWBs, and the only approach able to resolve the radio emission arising from the WCR. By studying the morphology of this region, we have been able to constrain the properties of the system like the wind-momentum rate ratio and the opening angle of the shock and, for the first time, indirectly estimate the positions of the two stars in the sky. Multi-epoch observations, combining very-high-resolution and multi-frequency data, spread along a decade ($\sim 10\%$ of the orbit) should allow us to trace a significant variation of these properties to better characterize Apep and the evolution of the two stars.

\section*{Acknowledgments}

The authors thank the anonymous referees for the insightful comments that have significantly improved the manuscript.
We also thank P.~A.~Crowther, I.~Negueruela, and P.~Williams for useful discussions, and C.~Day for helping during the observations.
The Long Baseline Array is part of the Australia Telescope National Facility which is funded by the Australian Government for operation as a National Facility managed by CSIRO. This paper includes archived data obtained through the Australia Telescope Online Archive (\url{http://atoa.atnf.csiro.au}). This work was supported by resources provided by the Pawsey Supercomputing Centre with funding from the Australian Government and the Government of Western Australia.
This work has made use of data from the European Space Agency (ESA) mission {\it Gaia} (\url{https://www.cosmos.esa.int/gaia}), processed by the {\it Gaia} Data Processing and Analysis Consortium (DPAC, \url{https://www.cosmos.esa.int/web/gaia/dpac/consortium}). Funding for the DPAC has been provided by national institutions, in particular the institutions participating in the {\it Gaia} Multilateral Agreement.
BM acknowledges support from the Spanish Ministerio de Econom\'ia y Competitividad (MINECO) under grant AYA2016-76012-C3-1-P and from the Spanish Ministerio de Ciencia e Innovaci\'on under grants PID2019-105510GB-C31 and CEX2019-000918-M of ICCUB (Unidad de Excelencia ``Mar\'ia de Maeztu'' 2020-2023).
JRC thanks the Nederlandse Organisatie voor Wetenschappelijk Onderzoek (NWO) for support via the Talent Programme Veni grant.
YH acknowledges the traditional owners of the land, the Gadigal people of the Eora Nation, on which the University of Sydney is built and some of this work was carried out.
This research made use of {\sc APLpy}, an open-source plotting package for {\sc Python} \citep{aplpy2012}; {\sc Astropy}, a community-developed core {\sc Python} package for Astronomy \citep{astropy2013}; and {\sc Matplotlib} \citep{hunter2007}.
This research made use of NASA’s Astrophysics Data System.

\bibliographystyle{mn2e}
\bibliography{bibliography.bib}

\appendix
\section{Contact Discontinuity from two spherically-symmetric winds} \label{sec:app}

As presented in the main paper, the bow-shaped radio emission observed in Apep is fully consistent with emission arising from a WCR.
Given that the shock is expected to be adiabatic due to the scales of the system \citep{stevens1992}, and that the two shock fronts remain unresolved in the LBA data (see Sect.~\ref{sec:radioVSoptical}), we would expect the contact discontinuity (CD) to be a plausible proxy to describe the morphology of the radio emission.

In this appendix we provide the detailed description of a CD in the ideal case of a binary system where the two stellar winds are spherically-symmetric and homogeneous, following the description done by \citet{canto1996}.
We note that this scenario does not properly characterise the environment of Apep, where an asymmetric wind may be expected, but its simplicity allowed us to provide some meaningful estimations for the system with a minimal number of assumptions.

\subsection{Derivation of the CD}

\begin{figure}
    \includegraphics[width=0.475\textwidth]{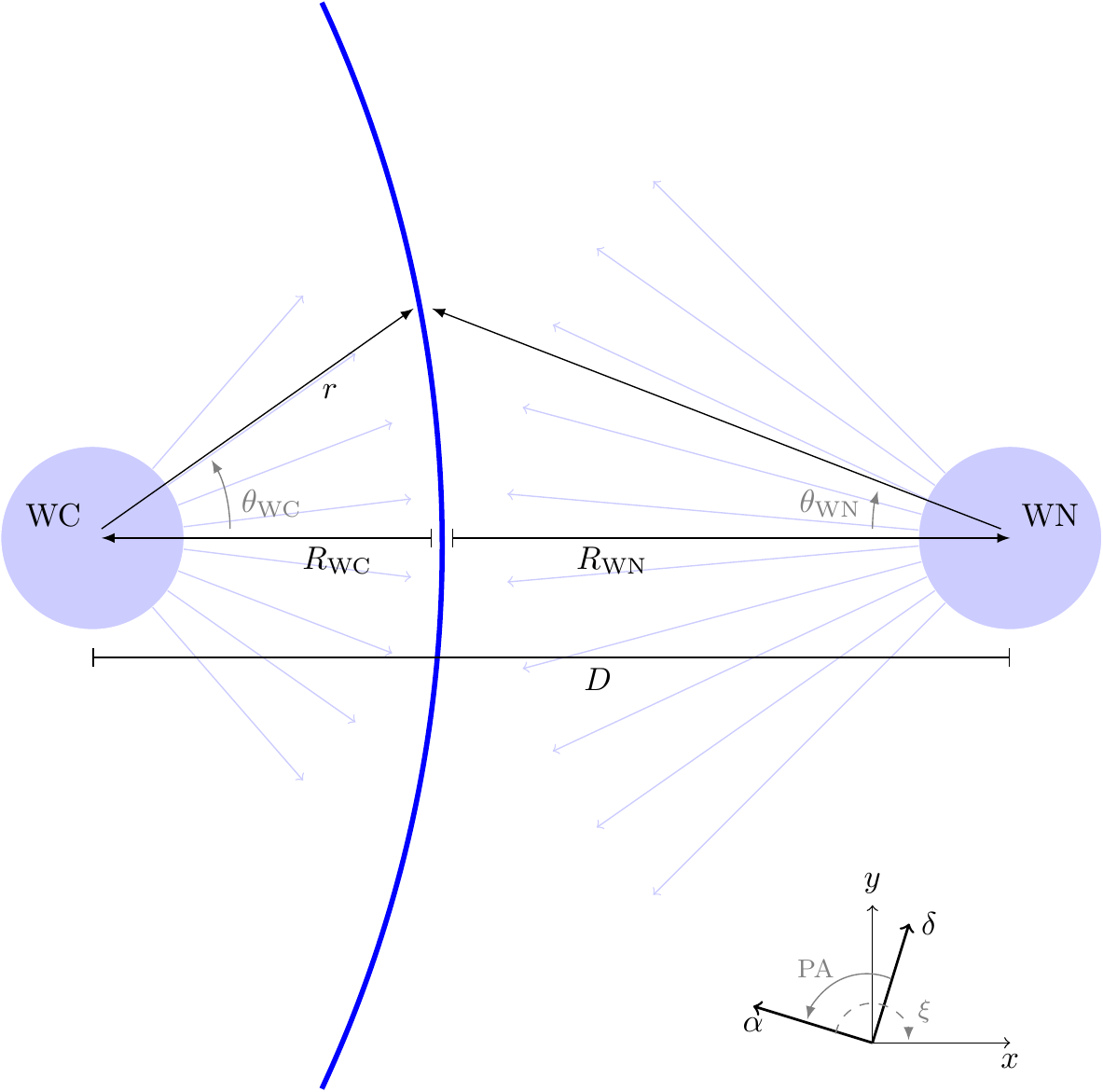}
    \caption{Schematic illustration of the contact discontinuity (CD) and the notation used in this study. The two stars are represented by the blue circles and the thick blue line represents the WCR, where the ram pressure of the two stellar winds (thin blue arrows) is balanced.}
    \label{fig:sketch}
\end{figure}
Under the assumption of spherically-symmetric stellar winds and adiabatic shocks, if these two shocks remain close enough so they are not resolved, then they would be expected to encompass the CD. Following \citet{canto1996}, the CD can be then characterized analytically as the interaction front where the ram pressure of the ideal stellar winds of the two stars (WC and WN in the case of Apep) are balanced (see Fig.~\ref{fig:sketch}): $\rho_{\rm WC} v_{\rm WC,\perp}^2 = \rho_{\rm WN} v_{\rm WN,\perp}^2$ (where $\rho$ and $v_{\perp}$ are the density and the wind velocity normal to the front, respectively). It has been shown \citep[following][]{canto1996} that this front can be parametrized as:
\begin{equation}
	r\, (\theta_{\rm WC}, \theta_{\rm WN}) = D \sin \theta_{\rm WN} \sin^{-1}(\theta_{\rm WC} + \theta_{\rm WN}), \label{eq:R}
\end{equation}
where $r$ is the separation of the front to the WC star for given $\theta_{\rm WC}$ and $\theta_{\rm WN}$ angles, and $D$ is the separation between the two stars. See Fig.~\ref{fig:sketch} for a representation of this front. The two angles are related by the expression
\begin{equation}
	\theta_{\rm WN}\, \tan^{-1} \theta_{\rm WN} = 1 + \eta \left( \theta_{\rm WC}\,\tan^{-1}\theta_{\rm WC} - 1 \right), \label{eq:thetas}
\end{equation}
where $\eta$ is the wind momentum rate ratio defined as:
\begin{equation}
	\eta \equiv \frac{\dot M_{\rm WC} v_{\rm WC}}{\dot M_{\rm WN} v_{\rm WN}} = \left( \frac{R_{\rm WC}}{R_{\rm WN}} \right)^2, \label{eq:eta}
\end{equation}
where $\dot M,\ v$, and $R$ are the mass-loss rate, stellar wind velocity, and distance to the stagnation point relative to the WC and WN stars, respectively. We note that $\eta$ is independent of the orbital inclination, and thus can be directly measured for any system once the positions of the two stars and the WCR are measured.
We remark that under the described scenario, the two stellar winds are assumed to be spherically-symmetric, and thus this value of $\eta$ is assumed to remain constant.

The asymptotic angle $\theta_{{\rm WC},\infty}$ of the bow shock (corresponding to $r \rightarrow \infty$) can be found from equation (\ref{eq:thetas}) given that both angles must verify $\theta_{\rm{WC},\infty} + \theta_{\rm{WN},\infty} = \pi$:
\begin{equation}
	\theta_{\rm{WC},\infty} - \tan \theta_{\rm{WC},\infty} = \pi\, ( 1 - \eta )^{-1}, \label{eq:tinf}
\end{equation}
which is related to the shock full opening angle $2\theta_w = 2(\pi - \theta_{\rm{WC},\infty})$. This angle can be estimated from the morphology typically observed in resolved radio-emitting WCRs, and can then be used to estimate the positions of the two stars.
Both the separation between the stars and the stagnation point can be directly obtained by following equations (\ref{eq:R}) and (\ref{eq:thetas}) to be:
\begin{equation}
	R_{\rm WC} = \frac{\eta^{1/2} D }{1 + \eta^{1/2}},\qquad
	R_{\rm WN} = \frac{D}{1 + \eta^{1/2}},\label{eq:posstar}
\end{equation}
where $R_{\rm WC}$ and $R_{\rm WN}$ are defined as in Fig.~\ref{fig:sketch}.

\subsection{Placing the CD in the plane of the sky}

Equations (\ref{eq:R}) and (\ref{eq:thetas}) define the CD region for the reference system shown in Fig.~\ref{fig:sketch}, where the origin of coordinates is placed at the center of the WC star and the $x$ axis in the direction of the WN star.
For systems where the orbital plane lies almost in the plane of the sky \citep[as in the case of Apep, with an inclination of $\sim \pm 25^\circ$;][]{han2020}, the radio emission would be seen as a conical structure where the front envelope can be parametrized by the following CD curve (in Cartesian coordinates):
\begin{align}
	x_{\rm CD} &= r\, \cos \theta_{\rm WC},\nonumber\\
	y_{\rm CD} &= r\, \sin \theta_{\rm WC};
\end{align}
and, by using the relations in equation (\ref{eq:R}):
\begin{align}
	x_{\rm CD} &= \frac{D}{2} \left[ 1 + \sin(\theta_{\rm WN} - \theta_{\rm WC}) \sin^{-1}(\theta_{\rm WC} + \theta_{\rm WN})\right],\nonumber\\
	y_{\rm CD} &= \frac{D}{2} \left[\cos(\theta_{\rm WN} - \theta_{\rm WC}) \sin^{-1}(\theta_{\rm WC} + \theta_{\rm WN}) - \tan^{-1}(\theta_{\rm WC} + \theta_{\rm WN})\right]. \label{eq:Axy}
\end{align}

In absence of significant absorption, as in the case of Apep, the radio emission is expected to be maximal at the stagnation point, and quickly decay for larger values of $\theta$. Under this scenario, and for a system like Apep (with an orbit on the plane of the sky), it can then be expected that the given envelope would gather the bulk of the radio emission following projection reasoning (e.g. such envelope would collect the highest values of the column density).

A generalized form for equation (\ref{eq:Axy}) is however needed given that the system can have a random orientation and is placed at some sky coordinates $(\alpha, \delta)$. Furthermore, the positions of the stars were {\em a-priori} unknown. Only the position of the radio-emitting WCR could be directly determined from the LBA data.
We therefore chose a better reference system where the origin of coordinates is placed at the stagnation point of the CD (which is expected to coincide with the peak of the reported radio emission, $\alpha_{\rm WCR}, \delta_{\rm WCR}$), and rotated by an angle $\xi$. We can then recover the CD curve in sky coordinates:
\begin{equation}
	\begin{pmatrix}
		\alpha_{\rm CD}\cos\delta_{\rm WCR}\\
		\delta_{\rm CD}
	\end{pmatrix} =
	\begin{pmatrix}
		\cos\xi & -\sin\xi\\
		\sin\xi & \cos\xi
	\end{pmatrix}
	\cdot
	\begin{pmatrix}
		x_{\rm CD}  - R_{\rm WC}\\
		y_{\rm CD}
	\end{pmatrix} +
	\begin{pmatrix}
		\alpha_{\rm WCR}\cos\delta_{\rm WCR}\\
		\delta_{\rm WCR}
	\end{pmatrix},
	\label{eq:Axy2}
\end{equation}
where $\alpha_{\rm CD}, \delta_{\rm CD}$ are functions of $\theta_{\rm WC}$ and $\theta_{\rm WN}$, but it can reduced to only $\theta_{\rm WC}$ by numerically solving equation (\ref{eq:thetas}). We note that this transformations assume that the declination can be considered constant for the full CD region (i.e. the system subtends a small angle in the sky). As it can be seen, the final curve only depends on the following parameters: the position of the stagnation point in the sky ($\alpha_{\rm WCR}, \delta_{\rm WCR}$), the position angle of the system ($\xi$, which is related to the PA of the system mentioned in the main text by $\xi = \frac{5}{2}\pi - \mathrm{PA}$), the separation between the two stars ($D$), and the wind-momentum rate ratio ($\eta$)\footnote{The code used to obtain the CD curve is part of the {\tt Binaries} package that can be publicly found at \url{https://github.com/bmarcote/binaries} under GPLv3 license.}.

\subsection{Comparison with WCR VLBI images}

As mentioned in Sect.~\ref{sec:results}, typical radio images are described as a collection of point-like components (so-called clean components; see Fig.~\ref{fig:lba}). Each component contains information of its position ($\alpha_i, \delta_i$) and flux ($S_i$). One can then use the positions of all clean components with positive flux to fit the expected CD curve.

To compare the expected ideal CD with respect to the observed morphology of the radio-emitting WCR in Apep we first fixed the parameters $D$ and PA \citep[as both are provided by][]{han2020}, and the position of the stagnation point to match the centroid of the radio image (see Sect.~\ref{sec:results}). Only the $eta$ parameter was then free. To compare which geometry best explained the observed curvature, we computed different curves for 500 different values of $\eta$ ranging 0.2--0.8, each of them covering angles of $|\theta_{\rm WC}| \leq 60^\circ$. These limit values were chosen by a manual inspection of Fig.~\ref{fig:fit}. From preliminary trials we guaranteed that the given range for $\eta$ contained any reasonable value that could reproduce the observed curvature, and the aforementioned values of $\theta_{\rm WC}$ were the ones covering the region with significant radio emission (i.e. where all clean components were located; see Fig.~\ref{fig:fit}).

The determination of the most plausible values of $\eta$ was performed by computing the $\chi^2$ values from the cumulative separations of each clean component position ($\alpha_i, \delta_i$) to the CD curve ($\alpha_{\rm CD}, \delta_{\rm CD}$).
Given that the CD curve (for a given $\eta$) is defined as a discrete set of positions ($\alpha_{\rm CD}^i, \delta_{\rm CD}^i$), the $\chi^2$ values were computed as
\begin{equation}
	\chi^2 (\eta) = \sum_j \min_i \left\{  \frac{\left((\alpha_j - \alpha_{\rm CD}^i) \cos \delta_{\rm WCR}\right)^2}{|\alpha_{\rm CD}^i - \alpha_{\rm WCR}|\cos\delta_{\rm WCR} }  +  \frac{\left(\delta_j - \delta_{\rm CD}^i\right)^2}{|\delta_{\rm CD}^i - \delta_{\rm WCR}|} \right\}.
\end{equation}
That is, we estimated the angular separation between each clean component and the closest point to the CD curve. The resulting $\chi^2$ is then determined following the Pearson generalized form \citep{pearson1900}, where the separations between the data and the expected positions are weighted by the expected positions, which are relative to the stagnation point.

We then proceeded to compute the $\chi^2$ probability density function (p.d.f.) to obtain the most probable value of $\eta$, which can be approximated under Gaussian assumptions to be $f \propto \exp(-\chi^2/2)$ \citep[e.g. follow \S 5.9 from][]{dagostini2003}. We determined the expected value and variance by numerical integration of:
\begin{align}
	\bar{\eta} &= \int \eta f(\eta)\, \mathrm{d}\eta\nonumber,\\
	\sigma_{\eta}^2 &= \int \left( \eta - \bar\eta\right)^2\, f(\eta)\, \mathrm{d}\eta.
\end{align}
As mentioned in the text, we obtained a value of $\eta = 0.44 \pm 0.08$ (where the uncertainty represents the standard deviation), which then allowed us to estimate the opening angle by numerically solving equation (\ref{eq:tinf}), and infer the positions of the two stars by combining equations (\ref{eq:posstar}) and (\ref{eq:Axy2}).

To confirm the robustness of this result, we also fit the data by weighting the contribution of each clean component by its flux, thus higher-flux components had a stronger contribution to the analysis. We did not observe significant differences in the final results with respect to the aforementioned approach and we recovered the same values of $\eta$.

Finally, even when the obtained results explain accurately the observed emission and produce consistent physical parameters for Apep, it is worth to remark the few caveats that underlie our model, specially when applying it to the case of Apep.
The main ones are obviously related to the assumed ideal conditions and ideal CD shape. We note that the described model assumes the radio emission to follow the ideal CD shape. The radio emission is expected to be confined between two shock fronts, and we assume both of them to be close enough to be well described by the CD (as mentioned in Sect.~\ref{sec:results}). This approximation has been widely used in previous studies of CWBs, showing consistent results \citep[see e.g.][]{blomme2010,benaglia2015}. Furthermore, Fig.~\ref{fig:fit} shows that the radio emission fits nicely the predicted CD curve, and thus we can be confident that such approximation is also accurate in the case of Apep for the given resolution that we achieved with the LBA data. We note that the strong radio emission from Apep (roughly one order of magnitude brighter than any other known non-thermal radio emitting CWB), makes the system an exceptional case as it allows us to obtain reliable VLBI images with a large signal-to-noise, minimizing the typical concerns on these kinds of analyses \citep{pittard2006}.

The derived opening angle, on the other hand, may be less reliable. The radio emission is likely to be produced at the region where the opening angles of the shocks have not reached yet their asymptotic values \citep[see e.g.][]{pittard2006}. And finally, the compared mass-loss rates are actually sensitive to the stellar wind velocities and the $\eta$ estimation. In addition, the possibility that a system like Apep shows anisotropic winds for at least one of the stars would produce -- in this model -- an average value of $\eta$ that smears such differences.

In any case, we note that in our model we did not consider neither the inclination of the system nor the brightness profile of the WCR. Whereas the former one is not expected to have a significant effect on the $\eta$ parameter (Apep is an almost face-on system, \citealt{han2020}, and the bow-shape curvature does not vary significantly for low inclination systems), the brightness profile could allow us to estimate the physical conditions at the WCR. However, this would not have any effect on the estimated wind momentum rate ratio and it would imply a significant number of additional assumptions and ideal conditions. We thus consider that such analysis is beyond the scope of this paper and a more realistic, (magneto-)hydrodynamical models, would be always preferable.

\label{lastpage}

\end{document}